\documentstyle[prd,aps,eqsecnum,preprint,psfig,floats]{revtex}

\def\a{\alpha}
\def\b{\beta}

\def\e{\epsilon}
\def\g{\gamma}
\def\d{\delta}
\def\p{\phi}
\def\vp{\varphi}

\def\l{\lambda}
\def\L{\Lambda}

\def\del{\partial}
\def\ha{\frac{1}{2}}
\def\psibar{\overline{\psi}}
\def\sla#1{#1\!\!\!/}
\def\be{\begin{equation}}
\def\ee{\end{equation}}
\begin{document}
\draft
\tighten

\preprint{\vbox{\hfill SMUHEP/94--29 \\
          \vbox{\hfill OSU--NT--94--10}
          \vbox{\vskip0.5in}
          }}

\title{ The Mandelstam-Leibbrandt Prescription in Light-Cone
Quantized Gauge Theories }

\author{ Gary McCartor }
\address{ Department of Physics, Southern Methodist University,
Dallas, TX 75275 }

\author{ David G. Robertson }
\address{ Department of Physics, The Ohio State University,
Columbus, OH 43210 }


\maketitle

\begin{abstract}
Quantization of gauge theories on characteristic surfaces and in the
light-cone gauge is discussed.  Implementation of the
Mandelstam-Leibbrandt prescription for the spurious singularity is
shown to require two distinct null planes, with independent degrees of
freedom initialized on each.  The relation of this theory to the usual
light-cone formulation of gauge field theory, using a single null
plane, is described.  A connection is established between this
formalism and a recently given operator solution to the Schwinger
model in the light-cone gauge.

\end{abstract}


\section{ Introduction }

Like the other axial gauges, the light-cone gauge ($A^+\equiv
A^0+A^3=0$)\footnote{We define light-cone coordinates $x^\pm\equiv
x^0\pm x^3$ and use latin indices $(i,j,\dots)$ to index the
transverse components $x_\perp=(x^1,x^2)$.  A contraction of
four-vectors decomposes as $A\cdot B =\ha(A^+B^- + A^-B^+) - A^i B^i,$
from which we infer the metric $g_{+-}=g_{-+}=\ha,$
$g^{11}=g^{22}=-1,$ with all other components vanishing.  Derivatives
are defined by $\del_\pm\equiv\del/\del x^\pm$, $\del_i\equiv\del/\del
x^i.$}
is plagued by the occurrence of a ``spurious'' singularity in the
gauge field propagator, which is related to the residual gauge freedom
(in this case transformations that do not depend on $x^-$).  For the
light-cone gauge a consistent interpretation of this singularity seems
to be via the Mandelstam-Leibbrandt (ML) prescription
\cite{mandelstam83,leibbrandt84}.  A large number of calculations
using this prescription have been performed and all give sensible
results, in agreement, where comparison is possible, with
covariant-gauge calculations.  For a good overview of these and
related issues see Ref.\cite{bns91}.

A derivation of the ML form of the propagator has been given in Ref.
\cite{bdls85}, in the framework of equal-time canonical quantization.
It has further been shown that gauge theories formulated in this way
are renormalizable\cite{bds87} (although some nonlocal counterterms
are required).  A central feature of this formalism is that one does
not reduce completely down to the physical (transverse) degrees of
freedom.  A longitudinal component of the gauge field is retained, and
a corresponding ghost field.  The Hilbert space of the theory thus
possesses an indefinite metric.  Selection of a physical subspace
results in the recovery of a positive-semidefinite metric and
Poincar\'e invariance.

Light-cone quantization of this theory was discussed in
Ref.\cite{mr94}, where it was shown that a second characteristic
surface, nowhere parallel to the conventional light-cone initial-value
surface $x^+=0,$ is needed to correctly recover the ML form of the
propagator.  The unphysical fields in the theory are initialized along
this other surface, and proper attention must be paid to the inclusion
of boundary contributions in the construction of conserved charges,
for example, the Poincar\'e generators.  Thus we do not have a
strictly ``Hamiltonian'' formalism, with all fields evolving from a
single initial-value surface.  This type of situation is quite
familiar from the treatment of massless fields quantized on
characteristic surfaces, particularly in two spacetime
dimensions\cite{mccartor88}.

The discussion of Ref.\cite{mr94} was for simplicity limited to free
fields.  This was sufficient for identifying the relevant degrees of
freedom and determining the propagator, which was the object of
primary interest.  The purpose of the present paper is to show how the
construction of Ref.\cite{mr94} is generalized to interacting theories
in a simple Abelian context.  We shall begin by reviewing the free
gauge field quantized on characteristic surfaces.  We then discuss the
simplest extension of this, quantum electrodynamics.  Most of the
features present in more complicated cases (e.g., non-Abelian gauge
theory) are already present in QED, and it further allows us to
discuss the treatment of Fermi fields in the simplest nontrivial
setting.  The case of two spacetime dimensions is instructive in that
it requires some special treatment, and furthermore is exactly
solvable for vanishing fermion mass.  In Sect. 3 we establish a
connection between the formalism presented here and an operator
solution to the Schwinger model in the light-cone gauge given recently
by Nakawaki\cite{nakawaki}.

As we shall see, the central problem is that of determining the
algebra of the field operators. It will prove to be quite difficult to
find a set of commutation relations that result in the Heisenberg
equations exactly reproducing the equations of motion.  Thus we shall
be unable to construct an interacting theory quantized on
characteristics that is precisely isomorphic to the theory described
in Ref.\cite{bdls85}.  It is possible, however, to construct a theory
that is equivalent to the full theory on the physical subspace.  A
field redefinition, which has essentially the form of a residual gauge
transformation, allows us to use simple (free-field) commutation
relations to achieve this.  The resulting theory is simply the
``naive'' light-cone theory tensored with the unphysical fields, which
are now decoupled.  They may therefore be discarded by invoking the
physical subspace condition. In this way we obtain a better
understanding of the relation between the ordinary light-cone
formulation of gauge theories and the formulation with the ML
prescription implemented.

\section{ Quantum Electrodynamics }

We begin by recalling the main results of Ref.\cite{mr94}.  The
light-cone gauge is parameterized by $n_\mu A^\mu=0$ with $n_+=1$ and
$n_-=n_\perp=0$.  For a free gauge field we consider the Lagrangian
\begin{equation}
{\cal L}=-{1\over4}F_{\mu\nu}F^{\mu\nu} - \lambda n_\mu A^{\mu}\; ,
\end{equation}
where $F_{\mu\nu}\equiv\del_\mu A_\nu - \del_\nu A_\mu$, and $\l$ is a
Lagrange multiplier field whose equation of motion enforces the gauge
condition $A^+=0.$ It satisfies
\begin{equation}
\del_-\l=0\; ,
					\label{lambdaeom}
\end{equation}
which indicates that it must be initialized along a surface of
constant $x^-,$ rather than the usual surface $x^+=0$.  It turns out
to be conjugate to a field $\vp,$ which is related to the value of the
transverse field $A^i$ at longitudinal infinity ($x^-=\pm\infty$).
Thus $\l$ and $\vp$ satisfy an equal-$x^-$ commutation relation, while
the remainder of the transverse field has the conventional equal-$x^+$
commutator.  Furthermore, the boundary contributions to the Poincar\'e
generators (more generally, to all conserved charges) must be retained
to correctly incorporate contributions from $\l$ and $\vp$.  The
Hilbert space of this theory has an indefinite metric and we must in
the end project onto a physical subspace, in a way familiar from
quantization in covariant gauges.  Specifically, we define physical
states to be those annihilated by the positive frequency part of $\l$:
\begin{equation}
\l^{(+)} |{\rm phys}\rangle =0\; ,
\end{equation}
that is, states between which $\l$ has vanishing matrix elements.
Maxwell's equations and the Poincar\'e algebra are obtained in matrix
elements between these states, which furthermore have nonnegative
norm.

Let us now consider QED, defined by the Lagrangian
\begin{equation}
{\cal L} = -{1\over4}F_{\mu\nu}F^{\mu\nu}
-\l n_\mu A^\mu +\psibar(i\sla{\del}-m)\psi -gJ_\mu A^\mu
					\label{qedlagrangian}
\end{equation}
with $J^\mu\equiv\psibar\g^\mu\psi.$ Our approach will be the same as
that in Ref.\cite{mr94}: use the equations of motion to identify the
degrees of freedom and where they must be initialized, and then
attempt to determine the field algebra by demanding that the
Heisenberg equations correctly reproduce the Euler-Lagrange equations.
Finally, we must check that projection onto a suitable physical
subspace can be carried out in a consistent way.

The equations of motion that follow from the Lagrangian
(\ref{qedlagrangian}) are, with $A^+=0$,
\begin{equation}
\del_-^2A^-+\del_-\del_iA^i=-\ha gJ^+\; ,
					\label{gauss}
\end{equation}
\begin{equation}
2\del_+\del_-A^--\del_\perp^2A^--2\del_+\del_iA^i=2\l+gJ^-\; ,
					\label{ampere}
\end{equation}
\begin{equation}
(4\del_+\del_--\del_\perp^2)
A^i+\del_i(\del_-A^-+\del_jA^j)=gJ^i\; ,
					\label{kleingordon}
\end{equation}
\begin{equation}
i\del_-\psi_-=\ha\left(-i\a^i\del_i+m\b-g\a^iA^i\right)\psi_+\; ,
					\label{diracminus}
\end{equation}
\begin{equation}
i\del_+\psi_+=\ha gA^-\psi_+
	+\ha\left(-i\a^i\del_i+m\b-g\a^iA^i\right)\psi_-\; ,
					\label{diracplus}
\end{equation}
where we have defined the standard light-cone spinor projections
$\psi_\pm\equiv\ha\g^0\g^\pm\psi,$ and $\a^i=\g^0\g^i$ and $\b=\g^0$
are the original Dirac matrices.  In addition the field $\l$ satisfies
Eq. (\ref{lambdaeom}) even in the presence of interactions.  The
easiest way to see this is to apply $\del_\nu$ to both sides of the
equation of motion
\begin{equation}
\del_\mu F^{\mu\nu} = n^\nu\l + gJ^\nu
\end{equation}
giving
\begin{equation}
2\del_-\l + g \del_\mu J^\mu = 0\; .
\end{equation}
We then note that $\del_\mu J^\mu$ vanishes by the Dirac
equation.\footnote{This assumes that we have defined the current
operator in such a way that it is not anomalous, as is normally
required for consistency.}

Now because $\l$ satisfies Eq. (\ref{lambdaeom}), it must be
initialized along a surface of constant $x^-,$ as in the free theory
\cite{mr94}.  Its ``conjugate momentum'' can be identified by
considering the light-cone Gauss' law, Eq. (\ref{gauss}).  It is
convenient to formally integrate this equation and express $A^-$ in
terms of ``zero mode'' degrees of freedom, that is, the integration
constants that arise.  We write
\begin{equation}
\del_-A^-+\del_iA^i = -g{1\over2\del_-}J^+
+\del_\perp^2\p(x^+,x_\perp)\; ,
					\label{gausssoln1}
\end{equation}
where $1/\del_-$ is some particular antiderivative and $\p$ is an
arbitrary function of $x^+$ and $x_\perp.$ As emphasized in
Ref.\cite{mr94}, $\p$ is part of the classical data required to
determine the general solution of the field equations, and so
corresponds to a degree of freedom in the quantum field theory.
Because it satisfies $\del_-\p=0$ by definition, it must also be
initialized on a surface of constant $x^-$.  It turns out to be
essentially conjugate to $\l,$ as in the free theory; this will be
shown in detail below.

Integrating Eq. (\ref{gausssoln1}) again results in
\begin{equation}
A^-=-{1\over\del_-}\del_iA^i - g{1\over2\del_-^2}J^+
+(\del_\perp^2\p)x^- + \g(x^+,x_\perp)\; ,
\label{gausssoln2}
\end{equation}
where $\g$ is another apparently arbitrary integration constant.  As
in the free theory, however, there will be a constraint relating the
three zero mode fields $\l$, $\p,$ and $\g,$ so that only two of
them are independent.  We shall here take $\l$ and $\p$ to be the
independent quantities, and $\g$ to be the determined one.

In addition to $\l$ and $\p,$ there are of course degrees of freedom
associated with the transverse fields $A^i$; we shall return to these
below.  For the moment let us discuss the degrees of freedom
associated with the Fermi field.  In the usual light-cone treatment we
observe that if $\psi_+$ is specified on $x^+=0,$ then Eq.
(\ref{diracminus}) is an equation of constraint that determines
$\psi_-.$ Thus the actual fermionic degrees of freedom are contained
in $\psi_+$.  This is known to be correct for the free massive Fermi
field.\footnote{In the massless case one can worry about modes with
$k^+=k_\perp=0$, which represent quanta propagating precisely along
the surface $x^+=0.$ They therefore cannot be initialized there, and
additional information must be given on another characteristic surface
to make the theory complete.  This is certainly important in two
spacetime dimensions, where the modes under discussion constitute half
of the theory.  In 3+1 dimensions these modes are a set of measure
zero and are conventionally neglected.  Some further discussion of
this problem will be given when we consider the transverse gauge
field.}
Regarding the question of whether or not to include an arbitrary
$x^-$-independent function in the solution of Eq.  (\ref{diracminus})
for $\psi_-$, we note that any solution of the free massive Dirac
equation that is independent of $x^-$ has infinite energy, and so is
presumably unphysical.  Solutions of this type have been discussed
recently\cite{lb93}, but they do not seem to be necessary in the
construction of the free theory.  We shall here assume that this holds
true for interacting fields as well, and treat the Fermi field in the
conventional way.  This includes taking the field to vanish at
longitudinal infinity when constructing conserved charges.  Again,
this is known to be the correct procedure in the free theory.

This boundary condition on $\psi$ insures that the current
$J^+=2\psi_+^\dagger\psi_+$ vanishes at longitudinal infinity.  It
will be necessary, however, to impose the further condition that $J^+$
have no zero mode, i.e., that
\be
\int^\infty_{-\infty} dx^- J^+ = 0\; .
					\label{condition}
\ee
This is necessary, for example, to insure that the Hamiltonian density
is integrable over $x^+=0,$ as we shall see below.  This condition on
$J^+$ also has implications for possible definitions of
$\del_-^{-(1,2)}J^+$ in, e.g., Eqs. (\ref{gausssoln1}) and
(\ref{gausssoln2}).  We might define, for example,
\be
{1\over\del_-}J^+(x^-) \equiv \ha\int^\infty_{-\infty}
dy^- \e(x^--y^-)J^+(y^-)
\ee
\be
\e(x)=\left\{\matrix{ 1&\qquad x>0\cr
                     -1&\qquad x<0\cr}\right.\; ,
\ee
which is the Cauchy principal value in momentum space.  Iteration of
this to obtain $\del_-^{-2}$ is in general ill-defined, however,
unless the integrand has no zero mode.  The condition
(\ref{condition}) insures that essentially any antiderivative may be
used consistently on $J^+$.

We may implement this condition on the current simply by coupling
$A^-$ to
\be
{J^+}^\prime\equiv J^+-\lim_{L\rightarrow\infty} {1\over
2L}\int_{-L}^L dx^- J^+\; ,
\label{jplusredef}
\ee
instead of to $J^+$ directly.  We do not yet understand all of the
implications of this redefinition on the structure of the theory.  It
resembles the choice of a restricted space of test functions for the
elementary fields in light-cone quantization [see the discussion
following Eq.  (2.26)], but applied to a composite object.  Since it
involves the definition of the current operator, its effects may be
expected to be depend on the particular regulator employed and the
methods used to define operator products.  It is thus difficult to
make general statements.\footnote{The necessity of the limiting
procedure in Eq. (\ref{jplusredef}), and indeed this entire
discussion, suggests that it may be natural to consider the theory
defined on a finite interval in $x^-$, with suitable boundary
conditions imposed on the fields.  This regulates the infrared
behavior and allows a rigorous discussion of these issues, up to the
treatment of UV divergences.  This type of approach is discussed in
Refs.\cite{kr94,soldati94}.}

Finally let us discuss the transverse fields $A^i$.  We begin by
observing that the field $\p$ is related to the value of $A^i$ on the
longitudinal boundaries, which we denote by $A^i_\infty$:
\be
A^i_\infty \equiv \lim_{x^-\rightarrow\pm\infty} A^i\; .
\ee
The most physically motivated way of seeing this is to consider the
classical expressions for the energy and momentum and demand that
these be finite.  We have
\begin{equation}
P^\mu=\ha\int_a dx^-d^2x_\perp T^{+\mu} +
	\ha\int_b dx^+d^2x_\perp T^{-\mu}\; ,
					\label{momenta}
\end{equation}
where $a$ is the usual light-cone initial value surface $x^+=0$ and
$b$ are the boundary wings (see Fig. 1).  The need to retain the
contributions from the boundary surfaces is in fact quite general, and
follows from insisting that the generators be the same as those we
would construct in equal-time quantization\cite{mccartor88}.  For the
energy-momentum tensor we may take the gauge-invariant form
\begin{equation}
T^{\mu\nu}=-F^{\mu\l}F^\nu_{~\l}-gJ^\mu A^\nu
+i\psibar\g^\mu\del^\nu\psi-g^{\mu\nu}{\cal L}
-\l n^\mu A^\nu\; .
					\label{emt}
\end{equation}
Now for the four-momentum (\ref{momenta}) to be finite, a necessary
condition is that those components of the field strength $F_{\mu\nu}$
that appear in the integral over the surface $x^+=0$ must vanish at
longitudinal infinity.  These are $F^{+-}$, $F^{+i}$, and $F^{ij}$, as
may be seen from an inspection of Eq. (\ref{emt}).  We consider first
the component
\begin{eqnarray}
\ha F^{+-} &=& \del_-A^-\nonumber \\
&=& -\del_iA^i -g{1\over2\del_-}J^+ +\del_\perp^2\p(x^+,x_\perp)\; ,
\end{eqnarray}
where we have made use of Eq. (\ref{gausssoln1}).  Requiring that this
vanish on the boundary wings results in
\be
\del_iA^i_\infty = \del_\perp^2\p
\ee
(recall that the current and its antiderivative are taken to
vanish on the boundaries).  Thus
\be
A^i_\infty = \del_i\p\; .
\ee
Not surprisingly, $A^i_\infty$ has the form of a pure gauge.  It
follows that $F^{+i}$ and $F^{ij}$ will also vanish on the boundary
surfaces.

\begin{figure}
\centerline{
\psfig{figure=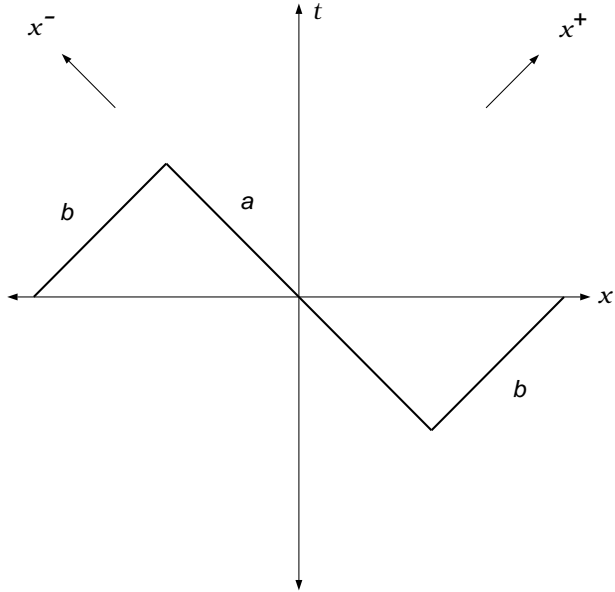,width=3.3in,height=4.3in}
}
\caption{Standard light-cone initial-value surface $x^+=0$ ($a$),
and boundary wings ($b$).}
\end{figure}

The fact that $A^i$ goes to a pure gauge at longitudinal infinity
suggests a natural field redefinition to disentangle the fields that
are initialized on $x^+=0$ from those that live on the boundaries.
Its form is simply that of a gauge transformation that removes the
boundary value of $A^i$.  Specifically we define
\begin{equation}
A^i\equiv T^i+\del_i\p\; .
					\label{aperpredef}
\end{equation}
It is consistent to take $T^i$ to vanish on the boundaries, as is
clear from the arguments presented above.  This can also be seen from
the equation of motion for the transverse fields.  Inserting Eq.
(\ref{gausssoln1}) into Eq.  (\ref{kleingordon}) we obtain
\begin{equation}
(4\del_+\del_--\del_\perp^2)A^i+\del_i\del_\perp^2\p
-g{1\over2\del_-}\del_iJ^+=gJ^i\; ,
					\label{kgmess}
\end{equation}
from which we see again that it is not consistent to assume that $A^i$
vanishes on the boundaries: the presence of the term
$\del_i\del_\perp^2\p$ means that even if $A^i$ is initially zero at
$x^-=\pm\infty$ it will not remain so under evolution in $x^+$.  In
terms of the redefined field, however, we have
\begin{equation}
(4\del_+\del_--\del_\perp^2)T^i-g{1\over2\del_-}\del_iJ^+=gJ^i\; .
					\label{tieom}
\end{equation}
It is thus consistent to assume that $T^i$ vanishes at longitudinal
infinity for all $x^+$.  As with massless fields generally, one can
consider the modes of $T^i$ that have $k^+=k_\perp=0.$ (Note that
these are a set of measure zero even relative to the fields $\l$ and
$\p.$) This is a familiar problem in the light-cone quantization of
massless fields, which is conventionally treated by choosing a test
function space in which to smear the field operators that has
vanishing support at the point $k^+=0$\cite{ss72}.  Thus they can be
consistently neglected, with the additional consequence that the
integral operator $1/\del_-$ will be well-defined when acting on
$T^i,$ since it has no $k^+=0$ modes.

Given $T^i$ and $\psi_+$ on $x^+=0,$ and $\p,$ $\l$, and $\g$ on the
boundary surfaces, we appear to have sufficient data to determine the
solution to Eqs. (\ref{diracplus}) and (\ref{tieom}) and solve for the
independent fields everywhere in spacetime.  All that remains is to
insure that this solution is consistent with Ampere's law, Eq.
(\ref{ampere}).  Substituting in the field redefinition
(\ref{aperpredef}), and making use of Eq.  (\ref{tieom}), we find that
Eq. (\ref{ampere}) reduces to a constraint relating the three zero
mode fields.  The precise form that this constraint takes depends on
the definition of the current, and in particular on the way the gauge
field becomes mixed with the Fermi field under renormalization.  This
will in general
depend on the regulator used.  Here we shall assume that the zero
mode fields do not mix with the current, as would presumably be true
for calculations using a gauge-invariant regulator such as
dimensional regularization.  In Sect. 3 we shall show
explicitly for the case of two dimensions that such mixing can happen
and that it modifies the details of the constraint for the zero mode
fields.  With the assumption of no mixing of the zero modes, however,
we obtain
\begin{equation}
\ha\g+\del_+\p+{1\over\del_\perp^2}\l = 0\; .
					\label{theconstraint}
\end{equation}
As noted previously, we shall take $\p$ and $\l$ to be the independent
quantities and $\g$ to be determined by Eq. (\ref{theconstraint}).

The final result of our analysis of the equations of motion is that
specifying $T^i$ and $\psi_+$ on $x^+=0,$ and $\p$ and $\l$ on the
boundary wings, is sufficient to determine a general solution to the
field equations (\ref{gauss})--(\ref{diracplus}).  These fields will
thus correspond to independent operators in the quantum theory.  The
algebra they satisfy may be determined by considering the Poincar\'e
generators, and requiring that the Heisenberg equations correctly
reproduce the field equations.

We focus on $P^\pm$, for which the relevant components of the
energy-momentum tensor (\ref{emt}) are
\begin{eqnarray}
T^{++} &=& 4(\del_-A^i)^2 + 4i\psi^\dagger_+\del_-\psi_+\; ,\\
\nonumber\\
T^{+-} &=& (\del_-A^-)^2+\ha(F^{ij})^2
+\psi_-^\dagger(-i\a^i\del_i+m\b-g\a^iA^i)\psi_++{\rm H.c.}\; ,\\
\nonumber\\
T^{--} &=& 4(\del_+A^i)^2+4(\del_+A^i)(\del_iA^-)+(\del_iA^-)^2-2\l A^-
+4i\psi^\dagger_-\del_+\psi_-\; ,\\
\nonumber\\
T^{-+} &=& (\del_-A^-)^2+\ha(F^{ij})^2-4i\psi^\dagger_+\del_+\psi_+
-2i\psi_+^\dagger\a^i\del_i\psi_--2i\psi_-^\dagger\a^i\del_i\psi_+
\nonumber \\
& &\qquad\qquad+2m(\psi^\dagger_-\b\psi_++ \psi^\dagger_+\b\psi_-)
+gJ^+A^--2gJ^iA^i\; .
\end{eqnarray}
We begin by expressing these, and the equations of motion
(\ref{diracplus}) and (\ref{tieom}), in terms of the independent
fields $T^i$, $\psi_+$, $\p$, and $\l$, keeping in mind that $T^i$ and
$\psi_\pm$ may be set to zero on the boundary surfaces.  The resulting
expressions are simplified considerably, however, by making the
further field redefinition
\begin{equation}
\psi_+\equiv e^{ig\p}\eta_+\; .
					\label{psiredef}
\end{equation}
For $A^i$ and $\psi_+$ the redefinitions (\ref{aperpredef}) and
(\ref{psiredef}) have the form of a residual ($x^-$-independent) gauge
transformation with the gauge function $\p$.\footnote{It is
interesting in this connection to examine the solution of Gauss' law
for $A^-$ in terms of the independent fields:
\begin{eqnarray}
A^-&=&-{1\over\del_-}\del_iT^i - g{1\over2\del_-^2}J^+-2\del_+\p
-{2\over\del_\perp^2}\l\nonumber\\
&\equiv& T^- -2\del_+\p-{2\over\del_\perp^2}\l\; .
\end{eqnarray}
Except for the term involving $\l,$ this can be interpreted as the
corresponding gauge transformation of $A^-$, where the ``transformed''
field $T^-$ is simply the naive expression for the constrained
component of the gauge field, written in terms of $T^i$ rather than
$A^i$.}
Note that the projection of the Dirac equation that determines
$\psi_-$ becomes, in terms of $\eta_+$ and $T^i$,
\begin{equation}
i\del_-\psi_-=e^{ig\p}\ha\left(
-i\a^i\del_i+m\b-g\a^iT^i\right)\eta_+\; .
\end{equation}
Thus the natural definition
\begin{equation}
\psi_-\equiv e^{ig\p}\eta_-
					\label{nature}
\end{equation}
results in $\eta_-$ satisfying the conventional constraint relation,
but written in terms of $\eta_+$ and $T^i$.  Furthermore, the currents
have the same form in terms of $\eta_\pm$ as they do in terms of
$\psi_\pm.$

In terms of the redefined fields we find
\begin{eqnarray}
T^{+-} &=& \eta_+^\dagger\left(-i\a^i\del_i+m\b-g\a^iT^i\right)
{1\over2i\del_-}
\left(-i\a^j\del_j+m\b-g\a^jT^j\right)\eta_+
+ {\rm H.c.}\nonumber \\
& &\qquad\qquad + (\del_iT^j)(\del_iT^j)+g(\del_iT^i)\Bigl({1\over\del_-}
J^+\Bigr) + {g^2\over4} \Bigl({1\over\del_-}
J^+\Bigr)^2\; ,
\end{eqnarray}
where we have dropped terms that will not contribute when we integrate
over $x^+=0$
to construct $P^-.$ The resulting contribution to $P^-$ is just the
naive light-cone Hamiltonian for QED, written in terms of $T^i$ rather
than $A^i.$  As anticipated, all coupling between the physical fields
$T^i$ and $\eta$ and
the zero modes has disappeared from $T^{+-}$.  Similarly
\begin{equation}
T^{++} = 4(\del_-T^i)^2 + 4i\eta^\dagger_+\del_-\eta_+\; .
\end{equation}
The components $T^{-+}$ and $T^{--}$, which are needed only on the
boundary surfaces, are precisely the same as in the free theory.  We
find that
\begin{equation}
T^{--} = 4\l(\del_+\p)
					\label{tmmonbdy}
\end{equation}
and
\begin{equation}
T^{-+} = 0
\end{equation}
on the boundaries.

The equations of motion are also easily rewritten in terms of the
redefined fields.  The solution to Gauss' law is
\begin{equation}
A^-=-{1\over\del_-}\del_iT^i - g{1\over2\del_-^2}J^+
-2\del_+\p-2\left({1\over\del_\perp^2}\l\right) \; ,
\end{equation}
and the equation of motion for the transverse field $T^i$ is Eq.
(\ref{tieom}).  Finally, the dynamical part of the Dirac equation [Eq.
(\ref{diracplus})] becomes
\begin{eqnarray}
i\del_+\psi_+&=&\ha\left(-i\a^i\del_i+m\b-g\a^iT^i
\right){1\over2i\del_-}
\left(-i\a^i\del_i+m\b-g\a^iT^i\right)\eta_+
\nonumber \\
& &\qquad\qquad-\ha g\left({1\over\del_-}\del_iT^i\right)\eta_+
-g\left({1\over\del_\perp^2}\l\right)\eta_+
-{1\over4}g^2\left({1\over\del_-^2}J^+\right)\eta_+\; .
					\label{redefdiracp}
\end{eqnarray}
Notice that the only difference between Eq. (\ref{redefdiracp}) and
what we would obtain in a naive light-cone quantization of QED is the
term involving $\l.$

We can now attempt to determine the commutation relations among the
fields such that Eqs. (\ref{tieom}) and (\ref{redefdiracp}) are
obtained in appropriate commutators of $T^i$ or $\eta_+$ with $P^-$.
In addition, they must give the correct results for commutators with
the ``kinematical'' generators, for example $P^+$, which translate the
fields within their respective initial-value surfaces. The free-field
commutation relations\cite{mr94}
\begin{eqnarray}
\left[T^i(x^-,x_\perp),\del^+T^j(y^-,y_\perp)\right]_{x^+=0}
&=& i\d^{ij}\d(x^--y^-)\d^{(2)}(x_\perp-y_\perp)
\label{tcommutator} \\ \nonumber\\
\left[\p(x^+,x_\perp),\l(y^+,y_\perp)\right]
&=& i\d(x^+-y^+)\d^{(2)}(x_\perp-y_\perp)
\label{plcommutator} \\ \nonumber\\
\left\{\eta_+(x^-,x_\perp),\eta_+^\dagger(y^-,y_\perp)\right\}
&=& \L_+\d(x^--y^-)\d^{(2)}(x_\perp-y_\perp)
\label{psicommutator}\\ \nonumber\\
\left[T_i,\p\right] = [T_i,\l] = [\p,\p] = [\l,\l] &=&
[\eta_+,T]=[\eta_+,\l]=\{\eta_+,\eta_+\} = 0
\label{misccommutator}
\end{eqnarray}
fulfill almost all of these requirements.  It is straightforward to
check that they give the correct Heisenberg equations, except that the
commutator of $\eta_+$ with $P^-$ fails to reproduce the term
containing $\l$ in Eq. (\ref{redefdiracp}). All the other Heisenberg
equations, including the kinematical ones, work correctly.

In order to obtain an interacting light-cone theory that is isomorphic
to the theory described in Ref.\cite{bdls85}, then, we would need to
impose commutation relations more complicated than
(\ref{tcommutator})--(\ref{misccommutator}).  Determining the
necessary field algebra would seem to be quite difficult.  The
required commutation relations would have to correctly give the term
proportional to $\l$ in Eq. (\ref{redefdiracp}), without of course
upsetting any of the other Heisenberg equations.  The Poincar\'e
algebra must furthermore be satisfied, up to terms proportional to
$\l$ (more on this below).  The required commutators could be computed
perturbatively, for example following the approach of Ref.\cite{rm92}.
One would quantize the theory at equal time and use perturbation
theory to solve for the fields everywhere in spacetime.  The resulting
fields could then be evaluated on the various surfaces of interest and
their algebra determined by direct computation.  Some work in this
direction is in progress.  Alternatively one could experiment with
further field redefinitions in an effort to obtain fields with simpler
commutators.

Let us turn for the moment to the issue of the projection onto a
physical subspace.  This subspace is defined by the requirement that
the equations of motion reduce to Maxwell's equations between physical
states.  Now, the only equation of motion that is not one of Maxwell's
equations is Eq. (\ref{ampere}), due to the appearance of $\l$ on its
right hand side.  Therefore $\l$ must have vanishing matrix elements
between physical states, or equivalently
\begin{equation}
\l^{(+)} |{\rm phys}\rangle = 0\; .
					\label{physicality}
\end{equation}
Note that in this subspace the extra term on the right side of Eq.
(\ref{redefdiracp}) vanishes.  Thus the theory defined by the
generators we have constructed with the free-field commutation
relations (\ref{tcommutator})--(\ref{misccommutator}) is equivalent to
the original theory on the physical subspace.  The resulting theory is
just the ``naive'' light-cone theory tensored with the unphysical
fields $\p$ and $\l$, which are decoupled.  They can therefore be
discarded by invoking the condition (\ref{physicality}). Clearly, the
states in the resulting theory have positive norm so that unitarity
holds, and furthermore the Poincar\'e algebra is satisfied.

Thus while it seems difficult to construct a light-cone version of the
theory (\ref{qedlagrangian}) that is isomorphic to the corresponding
equal-time theory, on the light cone there is a field redefinition
which essentially serves to disentangle the physical and unphysical
degrees of freedom.  This happens in such a way that simple
(free-field) commutation relations give a theory that is equivalent to
the full theory in the physical subspace.

\section{ The Schwinger Model }

We shall now show that for the case of the Schwinger model
(electrodynamics of massless fermions in two spacetime dimensions),
mixing of the gauge field with the Fermi field to form a renormalized
current operator modifies the details of the constraint equation for
the zero mode fields.  The general solution of the Maxwell equation
\be
\partial_-^2 A^- = -\ha g J^+\; ,
\ee
including the zero modes, is
\be
A^- = - g{1\over 2 \del_-^2}J^+ + \varphi(x^+) x^- + \g(x^+) \; .
					\label{amtd}
\ee
The main point we wish to make is that for the case of two dimensions
we know how the gauge field mixes with the Fermi field to form the
current.  If we use point-splitting and normal-ordering to define the
product of Fermi fields then we find\cite{mccartor91}
\be
J^- = j^-(x^+) -  {g \over 2 \pi}A^-\; ,
\ee
where
\be
j^- (x^+) \equiv \lim_{\epsilon\rightarrow0}
\left\{ \psi_-^\dagger(x^++\epsilon)\psi_-(x)
- {\rm V.E.V.}\right\}\; .
\ee
That $j^-$ is a function of $x^+$ only can be seen from the Dirac
equation:
\be
\partial_- \psi_- = 0\; .
\ee
If we now insert Eq. (\ref{amtd}) into the other Maxwell equation
\be
2\partial_+\partial_-A^- = 2\lambda + g J^-
					\label{amltd}
\ee
we see that $\varphi$ must be zero, since otherwise the RHS of Eq.
(\ref{amltd}) would contain a term linear in $x^-$ while the LHS would
not.  We therefore obtain for the constraint
\be
\lambda = {g^2 \over 4 \pi} \g - \ha g j^-\; .
\ee
If we define a field $\sigma$ by the relation
\be
\g \equiv {4 \sqrt{\pi} \over g} \partial_+ \sigma
\ee
we find that consistency between the equations of motion and the
Heisenberg equations forces $\sigma$ to be a ghost field (i.e., it
creates and destroys negative-norm states).  Furthermore, from the
Dirac equation
\be
i \partial_+ \psi_+ = \ha g A^- \psi_+
\ee
we find that we can write
\be
\psi_+ = e^{-i2\sqrt{\pi}\sigma}\eta_+\; ,
\ee
which is somewhat like Eq. (\ref{psiredef}), although the details
differ.

What we have found here is precisely the structure found in a full
light-cone gauge operator solution to the Schwinger model given by Y.
Nakawaki\cite{nakawaki}.
While it may appear to be possible to make a residual
($x^-$-independent) gauge transformation which would remove the field
$\sigma$ from both the Fermi field and the gauge field, that
possibility does not actually exist.  If we make the required
(nonlocal) gauge transformation, then the operator products necessary
to define the current as a split and gauge-corrected object become
undefined and the solution is destroyed.  Thus, while the operator
mixing changes the details, the situation in the Schwinger model is
much like in four dimensional QED: we must retain certain zero mode
fields, some of which must be ghosts. In both cases the purpose of
these fields is the same: to correct the operator products.

\section{ Discussion }

We have shown how to set up the classical boundary-value problem for
the theory (\ref{qedlagrangian}) defined on lightlike surfaces.  This
includes identifying the independent data, which correspond to
operators in the associated quantum field theory, and uncovering
constraints that follow from the equations of motion and from
requiring finiteness of the classical energy-momentum.  The problem we
have not solved is that of determining the field algebra that results
in the Heisenberg equations reproducing the exact field equations of
the theory.  There is a field redefinition, however, that results in
the almost complete disentangling of the physical and unphysical
fields, and allows a theory with simple commutation relations to be
defined that is equivalent to the full theory on the physical
subspace.  This simpler theory is just the naive light-cone QED.

The restriction (\ref{condition}) on the current that couples to $A^-$
appears to be necessary in the (continuum) light cone approach.  Its
precise meaning, however, is bound up with issues of regularization
and renormalization of the current operators, so that without being
more specific about these it is impossible to make definite
statements.

The extension of this work to the non-Abelian case is at present
somewhat unclear.  By demanding finiteness of the energy-momentum, we
again find that the boundary value of the transverse fields is related
to the unphysical field $\p^a$ which occurs in the first integral of
the light-cone Gauss' law. This boundary value is again a pure gauge,
with a gauge function given as an infinite series in the coupling, and
serves to motivate a field redefinition to disentangle the physical
and unphysical fields.  All of this does in fact go through as
desired, and what results are expressions for the energy-momentum in
which the physical and unphysical fields are decoupled, and equations
of motion which are of the usual light-cone form but containing extra
terms proportional to $\l^a$\cite{dgrunpub}.  It therefore appears
possible to define a theory which is equivalent to the full theory on
the physical subspace, as we have discussed here for QED.  The present
difficulty is that the boundary contributions to the Poincar\'e
generators differ from their free-field forms, so that they must have
complicated commutation relations among themselves in order to obtain
the correct Heisenberg equations.  We are currently studying this
problem, and hope to report more completely on the non-Abelian case in
the near future.

\acknowledgments

\noindent
It is a pleasure to thank A. Bassetto, K. Hornbostel, Y. Nakawaki, G.
Nardelli, and R. Soldati for helpful discussions.  G.M. was supported
in part by the U.S.  Department of Energy under Grant No.
DE-FG05-92ER40722.  D.G.R. was supported by the National Science
Foundation under Grants Nos. PHY-9203145, PHY-9258270, and
PHY-9207889.



\begin{references}

\bibitem{mandelstam83}
S. Mandelstam, Nucl. Phys. {\bf B213}, 149 (1983).

\bibitem{leibbrandt84}
G. Leibbrandt, Phys. Rev. D {\bf 29}, 1699 (1984).

\bibitem{bns91}
A. Bassetto, G. Nardelli, and R. Soldati, {\em Yang-Mills Theories in
Algebraic Non-Covariant Gauges} (World Scientific, Singapore, 1991).

\bibitem{bdls85}
A. Bassetto, M. Dalbosco, I. Lazzizzera, and R. Soldati, Phys. Rev. D
{\bf 31}, 2012 (1985).

\bibitem{bds87}
A. Bassetto, M. Dalbosco, and R. Soldati, Phys. Rev. D {\bf 36}, 3138
(1987).

\bibitem{mr94}
G. McCartor and D.~G. Robertson, Z. Phys. C {\bf 62}, 349 (1994).

\bibitem{mccartor88}
G. McCartor, Z. Phys. C {\bf 41}, 271 (1988).

\bibitem{nakawaki}
Y. Nakawaki, private communication.  His construction is based in part
on a solution to the Schwinger model described in Prog. Theor. Phys.
{\bf 72}, 134 (1984).

\bibitem{lb93}
N.~E. Ligterink and B.~L.~G. Bakker, Amsterdam University preprint, 1993.

\bibitem{kr94}
A.~C. Kalloniatis and D.~G. Robertson, Phys. Rev. D {\bf 50}, 5262 (1994).

\bibitem{soldati94}
R. Soldati, University of Bologna preprint, {\tt hep-th/9410043} (1994).

\bibitem{ss72}
S. Schlieder and E. Seiler, Comm. Math. Phys. {\bf 25}, 62 (1972).

\bibitem{rm92}
D.~G. Robertson and G. McCartor, Z. Phys. C {\bf 53}, 661 (1992).

\bibitem{mccartor91}
G. McCartor, Z. Phys. C {\bf 52}, 611 (1991).

\bibitem{dgrunpub}
D.~G. Robertson, unpublished.

\end{references}
\end{document}